\documentclass{vgtc}                          

\ifpdf
  \pdfoutput=1\relax                   
  \usepackage{graphicx}                
  \DeclareGraphicsExtensions{.pdf,.png,.jpg,.jpeg} 
\else
  \usepackage{graphicx}                
  \DeclareGraphicsExtensions{.eps}     
\fi%

\graphicspath{{figures/}{pictures/}{images/}{./}} 

\usepackage{microtype}                 
\PassOptionsToPackage{warn}{textcomp}  
\usepackage{textcomp}                  
\usepackage{mathptmx}                  
\usepackage{times}                     
\usepackage{cite}                      
\usepackage{tabu}                      
\usepackage{booktabs}                  

\usepackage{enumitem}

\onlineid{0}

\vgtccategory{Research}

\vgtcinsertpkg

\definecolor{avaviolet}{HTML}{555091}
\definecolor{avagreen}{HTML}{1f6e56}
\definecolor{avaorange}{HTML}{b54e00}
\definecolor{avablue}{HTML}{89a5dd}
\definecolor{darkgreen}{HTML}{006400}

\newcommand{\acro}[1]{\textsc{\MakeLowercase{#1}}}

\usepackage{lineno}

\title{Guided Data Discovery in Interactive Visualizations via \\ Active Search}

\author{
Shayan Monadjemi\thanks{e-mail: monadjemi@wustl.edu}\\ %
        \scriptsize Washington University in St. Louis %
\and Sunwoo Ha\thanks{e-mail: sha@wustl.edu}\\ %
     \scriptsize Washington University in St. Louis %
\and Quan Nguyen \\
     \scriptsize Washington University in St. Louis %
\and Henry Chai \\
     \scriptsize {\phantom{xxx}Carnegie Mellon University\phantom{xxxx}} %
\and Roman Garnett \\
     \scriptsize Washington University in St. Louis %
\and Alvitta Ottley\thanks{e-mail: alvitta@wustl.edu}\\ %
     \scriptsize Washington University in St. Louis %
}

\abstract{
Recent advances in visual analytics have enabled us to learn from user interactions and uncover analytic goals.
These innovations set the foundation for actively guiding users during data exploration. 
Providing such guidance will become more critical as datasets grow in size and complexity, precluding exhaustive investigation.
Meanwhile, the machine learning community also struggles with datasets growing in size and complexity, precluding exhaustive labeling.
Active learning is a broad family of algorithms developed for actively guiding models during training.
We will consider the intersection of these analogous research thrusts.
First, we discuss the nuances of matching the choice of an active learning algorithm to the task at hand.
This is critical for performance, a fact we demonstrate in a simulation study.
We then present results of a user study for the particular task of data discovery guided by an active learning algorithm specifically designed for this task.
} 

\CCScatlist{
  \CCScatTwelve{Human-centered computing}{Visual analytics}{}{};
  \CCScatTwelve{Human-centered computing}{Empirical studies in visualization}{}{}
  \CCScatTwelve{Computing methodologies}{Active learning settings}{}{}
}

\begin{document}

\firstsection{Introduction}
\maketitle


The visual analytics community has made significant strides in developing systems that facilitate the interplay between humans and machines in exploratory data analysis and sensemaking \cite{keim2008visual, endert2012semantic, kim2019topicsifter, pmlr-v70-jiang17d, siddiqui2018feedback,crouser2012affordance,crouser2013balancing}. 
These advances have enabled us to learn from user interactions and uncover their analytic goals. Moreover, they have set up the foundation for creating visual analytic systems that guide users during data exploration. 
Providing such guidance will likely become more critical as datasets grow in size and complexity, overwhelming the limited screen real estate and human cognitive power.



The nuances of providing guidance varies with respect to factors including the visual analytic task at hand \cite{ceneda2016characterizing}.
Prior work has explored a set of visual analytic tasks and corresponding techniques for providing guidance. For example Lin et al.~\cite{lin2017rclens} proposed a visual analytic system to aid users in detecting anomalous clusters in data. Similarly, Kucher et al.~\cite{kucher2017active} presented a visual analytic system to aid users in building accurate stance classification models. Extending upon prior work, we focus on the common, yet tedious, task of discovery: sifting through a dataset and identifying valuable data points efficiently. 
For example, an intelligence analyst may spend substantial time reviewing documents while unraveling a terrorist attack plot, many of which may be irrelevant to the attack. Likewise, a scientist may test numerous molecules -- incurring high cost -- while searching for a new drug candidate, many of which may prove useless in a medical setting. In these situations, the goal is to guide the user to discover as many relevant data points as possible while alleviating information overload caused by irrelevant information.


To identify an algorithm for guiding users in the task of discovery, we examine the family of \emph{active learning} algorithms. 
Active learning algorithms strategically guide the model during training to fulfil an analytic objective. 
First, we compare and contrast a set of active learning algorithms and demonstrate that their effectiveness is task-dependent. 
We demonstrate that choosing an appropriate technique for a given visual analytic task is key to performance. In doing so, we focus our experiments on the \acro{VAST} challenge 2011 dataset which was designed to mimic a realistic national security scenario~\cite{scholtz2012reflection,cook2014vast,grinstein2011vast}. 
Once we identify active search as the leading active learning algorithm for the task of discovery, we design a visual analytic proof-of-concept tool and augment it with active search. In this prototype, the interactive visualization is the medium of communication between the algorithm and the user. 
The user inspects data points sequentially and determines their relevance to the task at hand.
Simultaneously, the algorithm translates the observed interactions into labels for the underlying models and then guides the user by recommending the most promising data points for investigation.

Although there has been a history of integrating similar \acro{ML} approaches into visual analytic tools, the impact of these technologies on human behavior and contributing factors to their effectiveness are open for investigation. In a preliminary attempt to address this gap, we conduct a crowd-sourced user study to investigate the impact of the active search algorithm in assisting users during visual data exploration and discovery. 
Using our proof-of-concept system, participants saw a visualization of geo-tagged microblogs, which provided information about the spread of various disease symptoms. The task was to assist authorities by searching through social media posts to identify individuals who may be impacted by the potential epidemic. We present our user study results, highlighting some of the promises and challenges of this guided data discovery framework. The dataset and analysis scripts are available on GitHub~\protect \footnote{ \url{https://github.com/washuvis/vis2022guideddiscovery}}.

A summary of our contributions is as follows:
\begin{itemize}[noitemsep, topsep=0pt]

    \item We present a comparison between three active learning algorithms and demonstrate that their performance is task-dependent. Selecting an appropriate algorithm that maps to a given visual analytic task is critical.
    
    
    \item We map the interactive data discovery task into an active learning algorithm (namely active search) and design an ecosystem where user interactions inform underlying models and the algorithm guides the user through visual clues.
    
    \item We present a user study to validate the effectiveness of the active search algorithm in the interactive data discovery workflow. Our results show  consistent speedup in discovery throughput and fewer distractions by irrelevant portions of the data.
    
    
    
\end{itemize}

\section{Background}

In this section, we provide a brief overview on visual analytic themes related to learning from user interactions and guiding users during interactive analysis with a focus on active learning algorithms. 

\subsection{Active Learning}

As datasets grew in size and complexity, the machine learning (\acro{ML}) community faced the challenge of gathering labeled datasets. With exhaustive labeling no longer being a viable option, \acro{ML} researchers developed the family of \emph{active learning} algorithms to guide the training process. 
Active learning traditionally refers to the idea of learning algorithms choosing their training data strategically by querying an oracle
in order to achieve better predictive performance \cite{settles2009active}. 
Over time, \acro{ML} researchers have expanded the active learning paradigm to include new optimization objectives beyond model accuracy. 
Some examples include the discovery of as many members of a given class as possible (i.e.\ \emph{active search}~\cite{garnett2012bayesian}) and the detection of anomalous categories (i.e.\ \emph{rare category detection}~\cite{he2009nearest}). Recently, the visual analytics community has utilized these advances to incorporate humans in the active learning process via interactive interfaces \cite{kucher2017active, lin2017rclens, kim2019topicsifter}. To the best of our knowledge, this work is the first to incorporate and evaluate the \emph{active search} algorithm in an interactive setting.


\subsection{Learning from User Interactions}

A significant body of work in the visual analytics community has sought to enable human--machine partnerships in which machines are informed by low-level user interactions with interactive visualizations \cite{battle2016dynamic, ottley2019follow, dabek2016grammar, brown2014finding, bian2020deepva,xu2020survey,ragan2015characterizing}. 
Here, the technique of \textit{semantic interaction} is key, where user interactions with visualization tools translate into observations for underlying models, integrating user knowledge in the analysis process and informing intelligent response by the visualization system \cite{endert2012semantic}. 
Semantic interactions have been used to learn expert knowledge \cite{brown2012dis}, improve visual projection \cite{iwata2013active}, improve text analysis models \cite{kim2019topicsifter, kucher2017active, sevastjanova2018mixed}, and mitigate selection bias~\cite{gotz2016adaptive,monadjemi2020competingmodels}. 
%
For example, 
Brown et al.\ \cite{brown2012dis} proposed Dis-Function, a technique to represent expert knowledge as a distance function which is interactively learned by drag/drop interactions with a 2\acro{D} visualization. In their work, the user was not guided by an active learning algorithm to decide which points to move. Such decisions fully relied on human judgement.
To gauge how human judgement compares with those of active learning algorithms, Bernard et al.~\cite{bernard2017comparing} designed a study where the goal was to select training data for a classifier. They show that a classifier trained on data  selected by humans can outperform one trained on data selected by an active learning algorithm. Their findings rely on the underlying assumption that a 2\acro{D} visualization of the data is available and separates underlying class distributions. In this work, we evaluate a complete cycle, where the underlying models learn from user interactions and an active learning algorithm guides the user in discovery.

\subsection{Providing Visual Analytic Guidance}

In addition to using interactions to inform machine learning models, researchers have also developed systems where machines take actions to guide users in the analytic process \cite{guo2019visual, guo2019visualizing, jin2020carepre, das2019gaggle,ceneda2020guide,perez2022typology}. 
Ceneda et al.~\cite{ceneda2016characterizing} have defined \emph{guidance} as a computer-assisted process aimed at resolving users' knowledge gap during an interactive session. 
For example, Kim et al.\ \cite{kim2019topicsifter} proposed TopicSifter, an interactive system with the primary purpose of building models with high recall on text documents. 
This is an example of a system which guides the user by recommending potentially relevant keywords to include in the search space. 
Another example is \acro{ALVA} by Kucher et al.\ \cite{kucher2017active} which trains a stance classification model by querying the human user for labels.  
Yet another example of incorporating active learning into visual analytic workflows is RCLens by Lin et al.~\cite{lin2017rclens}. Their technique utilizes the rare category detection algorithm~\cite{he2009nearest} to guide users in identifying anomalies. In the context of our work, we utilize an active learning algorithm to guide the user to discover relevant data points efficiently while avoiding irrelevant ones.

\section{Selecting an Active Learning Algorithm}


Our primary focus in this section is the \emph{strategy} for selecting data points and their effectiveness in accomplishing a given analytic task.
Given a set of unlabeled data points, active learning algorithms sequentially pick one (or more) data points according to a \emph{strategy}. The selected data points are then presented to an oracle to acquire their labels. This oracle may be a human user or an expensive procedure (e.g.\ laboratory experiments). The observed labels are then used to re-train the underlying models.

In the simplest case, the learning algorithm selects data points from the set of unlabeled data at random. This strategy, known as \emph{random search}, does not aim to optimize any particular objective and is often used as a baseline in \acro{ML} literature. 
When the analytic goal is to train an accurate model, one approach is to identify points with the highest uncertainty and acquire their labels. This strategy, known as \emph{uncertainty sampling}, prioritizes data points on the boundary of classes where the model may be most confused about the labels \cite{lewis1994heterogeneous}. 
In some application areas, the primary objective is not to train an accurate model. Instead, the analyst seeks to identify as many members of a valuable class as possible. For such scenarios, Garnett et al.~\cite{garnett2012bayesian} have proposed an active learning approach called \emph{active search}.
We note that there are more active learning algorithms available in the \acro{ML} literature.
We focus on these three for demonstration purposes.
To show that choosing the appropriate active learning strategy for a given analytic goal is critical, we conduct simulations using the \acro{VAST} challenge 2011 dataset (described in Sec.~\ref{ss:scenario_data}). 
These experiments do not aim to simulate user behavior, but rather evaluate algorithm behavior for a set of analytic objectives.

\begin{table}[!b]
    \centering
    \scriptsize
    \caption{Comparing three active learning strategies for three analytic goals. The evaluation metrics are described in Section \ref{ss:simulation_setup}.}
    \newcolumntype{C}{ @{}>{${}}c<{{}$}@{} }
    \begin{tabular}{l *3{rCl} }
    \toprule
         & \multicolumn{3}{c}{Discovery} & \multicolumn{3}{c}{Detection} & \multicolumn{3}{c}{Training} \\
         & \multicolumn{3}{c}{\tiny{\# of microblogs}} & \multicolumn{3}{c}{\tiny{\# of symptoms}} & \multicolumn{3}{c}{\tiny{\acro{ROC-AUC}}} \\
         \midrule
         Random Search & 3.3 & \pm & 0.43 & 3.3 & \pm & 0.43 & 0.62 & \pm & 0.01 \\
         Greedy Active Search & 196.2 & \pm & 13.41 & 19.8 & \pm & 1.23 & 0.82 & \pm & 0.01 \\
         Uncertainty Sampling & 63.8 & \pm & 5.00 & 17.8 & \pm & 1.26 & 0.84 & \pm & 0.01 \\
         \bottomrule
    \end{tabular}
    \label{tab:sim_results}
\end{table}

\subsection{Scenario and Dataset}
\label{ss:scenario_data}
In \acro{VAST} Challenge 2011, the fictitious city of Vastapolis is under a biochemical attack, initiating an epidemic with flu-like symptoms. The health officials have access to a large collection of geo-tagged tweet-like microblogs. They want to identify impacted individuals, impacted neighborhoods, and the symptoms from social media posts. In doing so, they need to identify as many illness-related microblogs as possible. 
We select a sample of $50,000$ microblogs where only $1.3\%$ of them contain illness related content. We consider the illness related microblogs to be \emph{relevant} and the remaining ones to be \emph{irrelevant}. We encode the microblogs into a numerical space using an off-the-shelf \emph{word2vec} model, and build a $k$-\acro{NN} binary classifier using the cosine similarity metric. This model provides us with a mechanism to reason about the relevance of an unlabeled microblog in light of past observations. We cross validated these design choices against a set of alternatives to ensure our model is a viable choice.

\subsection{Simulation Setup}
\label{ss:simulation_setup}

Starting with one illness-related data point as our training set, we make 250 queries according to each of the following active learning strategies/algorithms: random search, greedy active search, and uncertainty sampling. We repeat this experiment 100 times and report evaluation metrics corresponding to three analytic objectives:
\textbf{Discovery} (the number of illness-related microblogs identified within the 250 queries),
\textbf{Detection} (the number of unique symptoms detected within the 250 queries), and
\textbf{Training} (the \acro{ROC-AUC} score of a $k$-\acro{NN} binary classifier trained on the 250 queried data points).

\subsection{Simulation Results}

The outcomes of our simulations are shown in Table \ref{tab:sim_results}. It is evident that different active learning strategies are dominant for different analytic objectives. When the objective is to discover members of a valuable class efficiently, the active search strategy outperforms the other strategies significantly. On the other hand, when the objective is to train an accurate classifier, the uncertainty sampling strategy outperforms the other strategies. Our results paint a consistent picture that choosing an appropriate active learning strategy for a task at hand is critical for performance.

\section{Example Prototype and User Study}

\begin{figure}[!t]
    \centering
    \includegraphics[width=\linewidth]{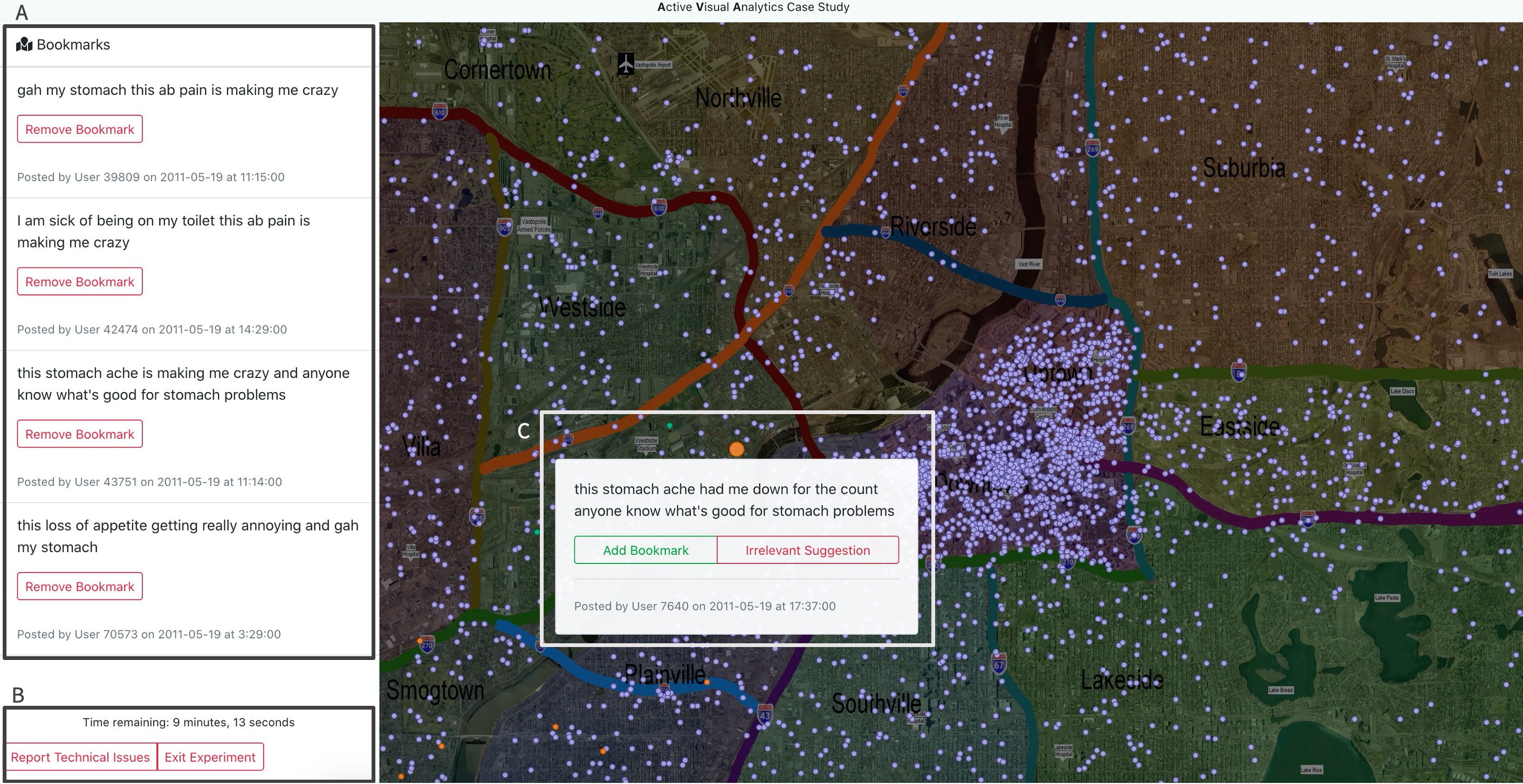}
\caption{A view of the prototype system for the epidemic dataset.}
    \label{fig:vastapolis_map}
\end{figure}

To investigate the impact of active search guidance on user behavior during visual exploration and discovery, we designed a crowd-sourced user study
{\protect \footnote{This experiment was pre-registered on \href{https://osf.io/6shr7/?view_only=389cd25b96f848a6a222d6966c82a325}{Open Science Foundation}.}}
that tasked participants to interact with a map of the fictional city of Vastapolis (Fig.~ \ref{fig:vastapolis_map}),
visualizing microblogs as dots placed on their posting locations. 
Hovering on data points triggered a tooltip containing the microblog and a bookmark button (Fig.~\ref{fig:vastapolis_map}, C). A countdown of the remaining time was shown, and users had the option to exit the experiment or report issues (Fig.~\ref{fig:vastapolis_map}, B). User bookmarks were listed in the side bar (Fig.~\ref{fig:vastapolis_map}, A).


\subsection{Task}

Participants were told that health professionals had reported a spike in illnesses with flu-like symptoms, including  fever,  chills,  sweats,  nausea and vomiting, and diarrhea.   
We informed participants that the authorities are interested in identifying the impacted parts of the city by analyzing social media activity. 
Their task was to assist the authorities by searching through a dataset of geo-tagged microblogs using the interactive interface shown in Figure~\ref{fig:vastapolis_map} and bookmarking as many posts containing illness-related information as possible. 
For this study, we narrowed the dataset to a sample of 3000 points from the approximate start of the epidemic, where $33\%$ of the data points contained illness-related content. We made these design choices to avoid an overly-crowded map and ensure that the discovery task is not prohibitively difficult for the participants.

\subsection{Participants}
We recruited 130 participants via Amazon's Mechanical Turk platform. Participants were 18 to 65 years old, from the United States, and fluent in English. Each participant had a \acro{HIT} approval rating of greater than 98\% with more than 100 approved \acro{HIT}s.
After data cleaning steps outlined in Section \ref{ss:data_collection_1}, 
there were 46 women, 76 men, and 1 participant with undisclosed sex in our subject pool with ages ranging from 18 to 62 years ($\mu = 36$, $\sigma = 9$). 72\% of our participants self-reported to have at least an associate degree. The average completion time (including reading the tutorial, performing the task, and completing the survey) was 12 minutes.
The instructions specified that participants will be compensated \$1.00 base pay and an additional \$0.10 bonus for every relevant microblog they identify (with a maximum of \$4.00). Although the advertised payment structure was designed to incentivize participants to complete the task, we paid everyone the maximum bonus of \$4.00 for fairness. 

\subsection{Procedure} \label{ss:procedures_1}
The experiment complied with an approved protocol per 
Washington University's \acro{IRB}. Workers who accepted the \acro{HIT} followed a \acro{URL} to the study platform. Our system randomly assigned each participant to one of the following groups: the \textit{active search group,} which received a batch of 10 active search queries in the form of visual clues (orange dots on the map, as shown in Fig.~\ref{fig:vastapolis_map}) that were updated after every bookmark interaction, and the \textit{control group,} which did not receive any assistance during exploration. 
Upon giving consent to participate in our study, participants were given a tutorial on their task and their corresponding system.  Both groups initiated their task without any initial ``clues,'' and in particular the active search group did not receive assistance for selecting their first bookmark.
Participants were given at most 10 minutes to identify as many microblogs related to the epidemic as they could using the interactive tool. 
Once the users were either satisfied with their search or the 10 minutes were up, they were directed to a post-experiment survey to collect demographic data, feedback on the system, and they self-reported their trust towards the suggestions on a Likert scale. In case our participants experienced technical difficulties, we provided an option to report issues, exit the session, and receive compensation.

\subsection{Data Collection, Cleaning, and Exclusions} \label{ss:data_collection_1}
We analyze our user study data by focusing on two interactions: inspection of microblogs (\textit{hovers)} and discovery of relevant ones (\textit{bookmarks}). These two types of interactions inform us about the speed and accuracy of visual data discovery through the metrics listed in Table \ref{tab:user_study}. Throughout this analysis, we consider illness-related microblogs to be \emph{relevant}. The bookmark and hover purity metrics are the proportion of all bookmarks and all hovers that involved relevant data points, respectively. The bookmarks- and hovers-per-minute metrics inform us about the speed at which users interacted with data points. The relevant hovers and relevant bookmarks-per-minute metrics are the rate at which users interacted with relevant data points, quantifying both speed and accuracy. Finally, we measure the number of relevant microblogs discovered by the end of the session and the number of unique symptoms in the discovered microblogs.

In a pre-processing step, we filtered the collected data to exclude participants who did not attempt the task or were unable to finish the experiment. Specifically, we eliminated participants based on the following four criterion:
(1) failed the survey attention checks (1 subject),
(2) reported technical issues (1 subject),
(3) hovered\footnote{We consider a valid hover to be one that lasts at least 500ms (300ms for triggering the tooltip, and 200ms for skimming the text)} on fewer than 10 data points (4 subjects), and
(4) did not meet the age qualification (1 subject).
%
A total of 123 subjects remained after filtering (74 in the \textit{control} group and 49 in the \textit{active search} group).





\subsection{Results}

\begin{table*}[t]
\centering
\scriptsize
\caption{The results of two-sample $t$-tests on the metrics discussed in Section \ref{ss:data_collection_1}}
\label{tab:user_study}
\begin{tabular*}{0.9\textwidth}{@{\extracolsep{\stretch{1}}}{l}*{5}{r}@{}}
\toprule
 & \multicolumn{2}{c}{95\% \acro{CI}} \\
 \cmidrule(lr){2-3} 
Metric & Control \tiny{N=74} & Active Search \tiny{N=40} & $p$-value & $t$-statistic & Cohen's $d$
\\ \midrule
Hovers per Minute & $16.7 \pm 1.19$ & $14.3 \pm 1.23$ & $\mathbf{0.0112}$ & $-2.58$ & $-0.51$ \\
Relevant Hovers per Minute & $6.7 \pm 0.68$ & $9.2 \pm 1.12$ & $\mathbf{0.0001}$ & 4.00 & 0.79 \\
Hover Purity & $0.39 \pm 0.02$ & $0.63 \pm 0.05$ & $\mathbf{< 0.0001}$ & 9.70 & 1.92 \\
\midrule
Bookmarks per Minute & $6.9 \pm 0.77$ & $9.5 \pm 1.41$ & $\mathbf{0.0006}$ & 3.52 & 0.70 \\
Relevant Bookmarks per Minute & $5.4 \pm 0.68$ & $8.1 \pm 1.26$ & $\mathbf{0.0001}$ & 3.98 & 0.79 \\
Bookmark Purity & $0.77 \pm 0.04$ & $0.82 \pm 0.05$ & $0.2249$ & 1.22 & 0.24 \\
\midrule
Relevant Microblogs Bookmarked & $53.9 \pm 6.80$ & $73.4 \pm 11.50$ & $\mathbf{0.0026}$ & 3.09 & 0.61 \\
Unique Keywords Identified & $16.1 \pm 0.83$ & $15.6 \pm 1.32$ & $0.4980$ & -0.68 & -0.13 \\
\bottomrule
\end{tabular*}
\vspace{-4mm}
\end{table*}


 
\noindent\textbf{Suggestion Quality: } We begin our analysis by examining the quality of suggestions provided by the active search algorithm when seeded with users' interaction data. We define suggestion purity to be the proportion of unique \emph{relevant} microblogs recommended to the user throughout a given session. On average, active search group participants had a suggestion purity of $79\%$. We observe a moderate positive correlation between bookmark purity and suggestion purity ($R^2_{adj}= 0.594$, $p < 0.0001$), suggesting that the active search algorithm provides useful recommendations for participants who interacted with microblogs containing known symptoms.

\begin{figure}[!ht]
    \centering
    \includegraphics[width=0.8\linewidth]{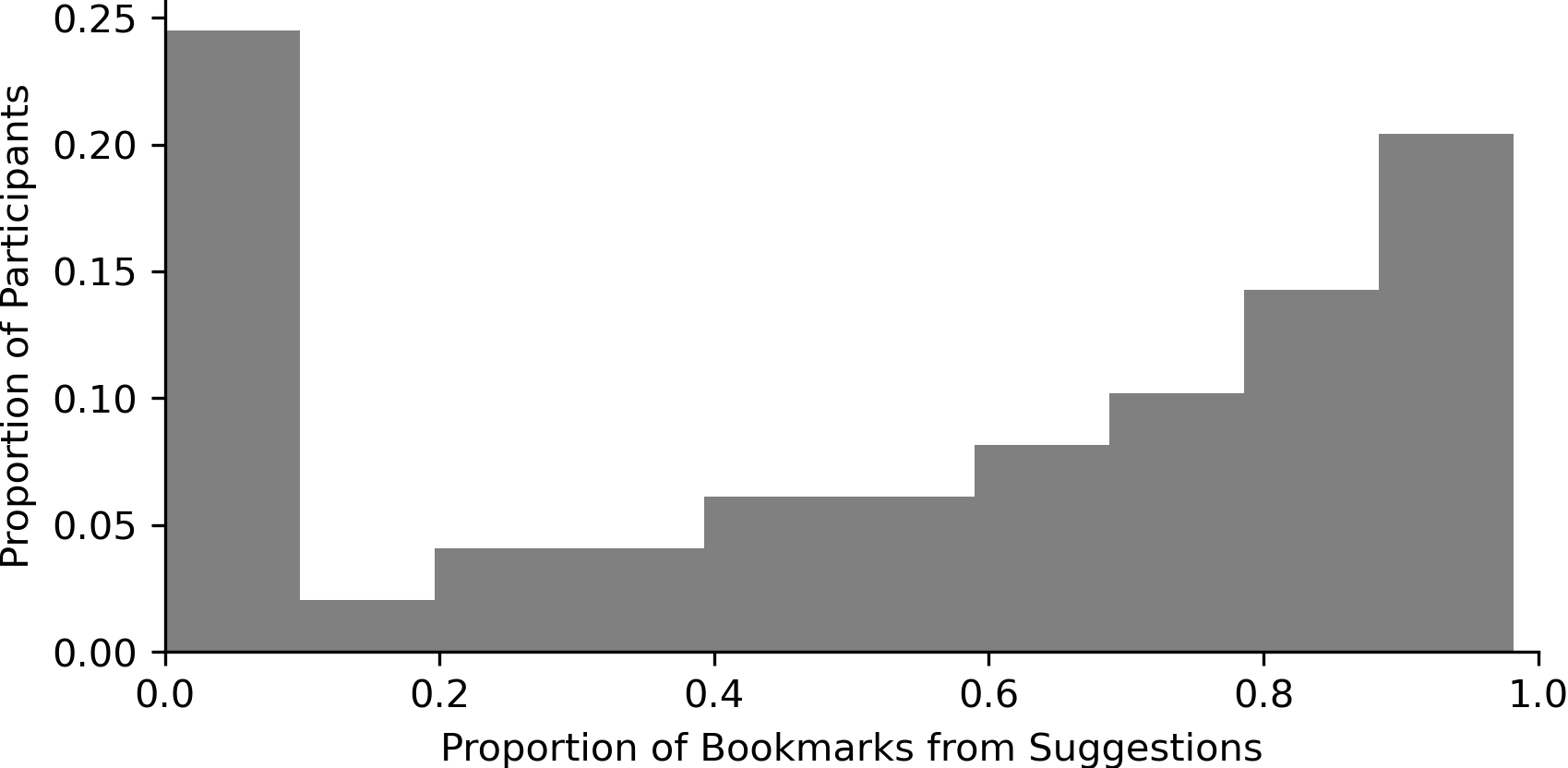}
     \caption{Distribution of proportion of bookmarks resulting from the active search suggestions.}
    \label{fig:rec_usage_pct_1}
    \vspace{-2mm}
\end{figure}

\vspace{.5em}
\noindent \textbf{Suggestion Usage: }
We observed an unexpected pattern in the active search group. As shown in Figure \ref{fig:rec_usage_pct_1}, for approximately $24\%$ of participants in the active search group, suggested microblogs accounted for less than 10\% of their bookmarks. 9 out of 49 participants did not bookmark any of the suggestions presented to them at all. Further inspection reveals that the 9 active search participants who ignored the suggestions had on average $82 \pm 9\%$ suggestion purity and $76 \pm 11 \%$ bookmark purity.
This compares to the the 40 active search participants who interacted with the suggestions, who had on average $79 \pm 5 \%$ suggestion purity and $82 \pm 5 \%$ bookmark purity.
Finally, we observed a difference between how subjects reported their trust towards system suggestions on a 1--5 Likert scale in the post-experiment survey ($3.3 \pm 0.46$ for those who ignored suggestions vs.\ $4.2 \pm 0.24$ for those who interacted with the suggestions). 


Moving forward, our analyses focus on the impact of suggestions on data exploration and discovery. Thus, we exclude the 9 participants in the \textit{active search} group who did not interact with the suggestions, leaving us with 74 participants in \textit{control} group and 40 participants in the \textit{active search} group. This additional filtering step further cleans our data to compare participants who had access to and used system recommendations to those who did not have access to recommendations. This additional filtering step does \emph{not} impact the conclusivity of our results in Table \ref{tab:user_study} (see the supplemental material).

\begin{figure}[!hb]
    \centering
    \includegraphics[width=0.8\linewidth]{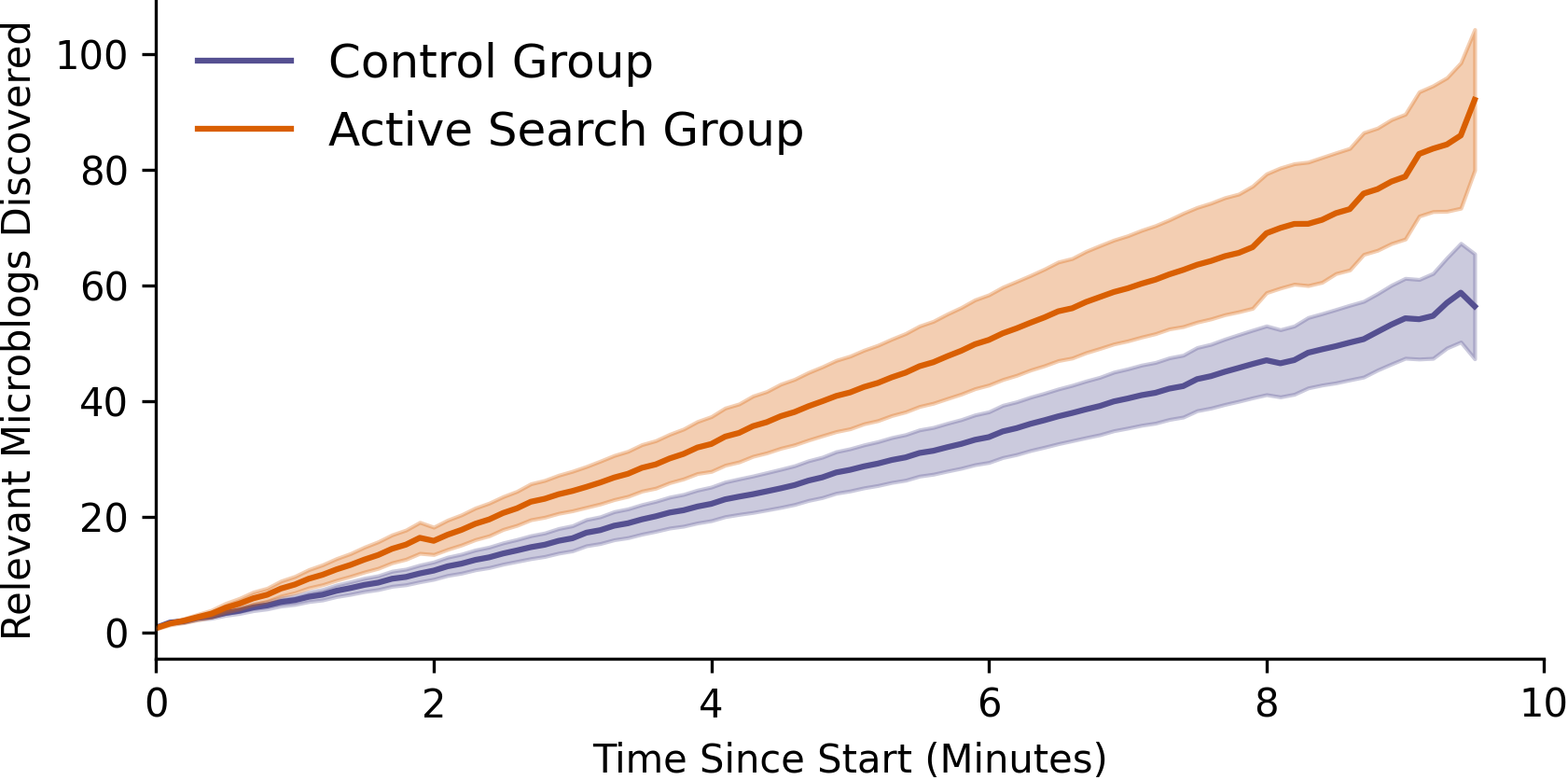}
    \caption{The average number of relevant microblogs discovered and the $95\%$ confidence interval for each group.}
    \label{fig:bmovertime1}
\end{figure}

\vspace{.5em}
\noindent\textbf{The Effect of Suggestions on Data Discovery: }
We performed a series of two-sample $t$-tests to investigate differences in behavior in our two study conditions: \textit{control} and \textit{active search}. Table \ref{tab:user_study} summarizes our findings. We found that participants in the \textit{active search} group bookmarked 
and hovered over 
significantly more relevant microblogs per minute than the \textit{control} group.  
Furthermore, our findings show that the \textit{active search} group performed fewer exploratory hovers per minute 
than the \textit{control} group, implying that the suggestions resulted in a more efficient exploratory analysis.  
For a more fine-grained analysis, we examine bookmark discoveries as a function of time. Figure~\ref{fig:bmovertime1} shows the average number of bookmarks over time for the \textit{active search} and \textit{control} groups. We observe that the \textit{active search} group consistently outperformed the control group by bookmarking more relevant microblogs throughout the ten-minute session. However, it is noteworthy that although the suggestions improved the quantity of the bookmarks, we found no measurable difference in the quality or content of the bookmarked discoveries. Both the \textit{active search} and \textit{control} groups collectively examined similar geographical regions and symptom sets.

\vspace{.5em}
\noindent \textbf{Impact of Active Search Suggestions on Usability: }
In a post-experiment survey, we asked subjects in both groups three questions on \textit{willingness to use}, \textit{ease of use}, and \textit{ease of task completion}. We performed a Mann--Whitney U statistical test at $\alpha = 0.05$ to determine if there was a significant difference between the control and active search groups. The analysis showed some evidence that the active search group found the system easier to use ($U = 1199.50$, $p = 0.0336$, $r = 0.16$) and were more willing to use the system frequently ($U = 1192$, $p = 0.0362$, $r = 0.16$). However, the effect sizes were small for both. Furthermore, we did not find a significant difference between the control and active group's response to ease of task completion ($U = 1397.50$, $p = 0.2989$, $r = 0.07$).

\section{Discussion and Conclusions}

We considered the task of discovery: identifying relevant data points to the task at hand efficiently. 
Upon examining the family of active learning algorithms, we selected active search as the leading technique for guiding this discovery process. 
Our user study results validate that guidance provided by the active search algorithm can significantly improve interactive data discovery. 
Our findings show that the participants successfully disregarded \emph{irrelevant} information and were more mindful towards the \emph{relevant} data points. These results can have high-impact implications for designing visual analytic tools for tasks such as intelligence analysis and scientific discovery.
Unexpectedly, we observed a nontrivial number of participants who ignored active search recommendations despite their relevance to their task. This highlights an avenue for investigating how to present such recommendations effectively in similar visual analytic systems. 



\acknowledgments{
This work is supported in part by the National Science Foundation under Grant No. OAC-2118201, OAC-1940224, and IIS-1845434.}

\bibliographystyle{abbrv-doi}

\bibliography{paper}
\end{document}